\documentclass[aps,prl,twocolumn,groupedaddress,showpacs,amssymb,amsmath]{revtex4}
\usepackage{dcolumn}
\usepackage{graphicx}

\begin{document}

%\preprint{}

\title{Magnetism in Disordered Graphene and Irradiated Graphite}

\author{Oleg V. Yazyev}
\email[]{oleg.yazyev@epfl.ch}
\affiliation{Ecole Polytechnique F\'ed\'erale de Lausanne (EPFL), Institute of Theoretical Physics (ITP) \\
and Institute of Chemical Sciences and Engineering (ISIC), CH-1015 Lausanne, Switzerland}
\affiliation{Institut Romand de Recherche Num\'erique en Physique
des Mat\'eriaux (IRRMA), CH-1015 Lausanne, Switzerland}

\date{\today}



\begin{abstract}
The magnetic properties of disordered graphene and irradiated graphite are systematically
studied using a combination of mean-field Hubbard model and first principles calculations. 
By considering large-scale disordered models of graphene I conclude that only single-atom 
defects can induce ferromagnetism in graphene-based materials. Preserved stacking 
order of graphene layers is shown to be another necessary condition for achieving a finite 
net magnetic moment of irradiated graphite. {\it Ab initio} calculations of hydrogen 
binding and diffusion and of interstitial-vacancy recombination further confirm the crucial role 
of stacking order in $\pi$-electron ferromagnetism of proton bombarded graphite.
\end{abstract}

\maketitle

% Introduction

Graphene and related nanostructured materials are recognized
as possible building blocks for prospective electronics technologies 
\cite{Avouris07} among which spintronics  \cite{Zutic04} occupy 
a special place. Extraordinary magnetic properties of carbon 
nanostructures resulting from their reduced dimensions have been predicted 
theoretically \cite{Fujita96-Son06a,Brey07,Fernandez-Rossier07}. 
This progress allowed to think of novel devices realizing 
spintronics in practice \cite{Son06b-Yazyev08a-Wang08}. 
Although most of the predicted properties are still awaiting their  
experimental confirmation, a discovery of room-temperature 
ferromagnetism in proton irradiated graphite \cite{Esquinazi03} 
continues to stimulate research in the field of magnetic carbon.

% Introduction 2
Recent  experimental investigations have revealed that the magnetic 
order in proton bombarded graphite has two-dimensional (i.e.~graphene-like) character \cite{Barzola-Quiquia07} 
and originates from the carbon $\pi$-electron system rather than
from intrinsic or introduced impurities \cite{Ohldag07}. These 
results are supported by theoretical calculations \cite{Yazyev07a,Kumazaki07} 
which show the crucial role of defect-induced quasilocalized states 
\cite{Mizes89-Kelly98,Pereira06} in producing a magnetic order stable 
at high temperatures. Although magnetic properties of defects in graphene
have been widely studied by first principles methods, the use of 
different periodic models of limited size often resulted in 
varying outcomes \cite{Lehtinen04,Duplock04,Yazyev07a,Pisani07b,Boukhvalov08}.
This calls for a study employing realistic disordered models. The origin 
of ferromagnetic order in graphene which generally favors antiferromagnetic coupling
between the neighboring atoms \cite{Yazyev07a,Brey07} as well as the role of 
various possible defects in inducing magnetism remain elusive.

% Contents
In this Letter, I address magnetic properties of disordered graphene 
and irradiated graphite by using a combination of mean-field Hubbard 
model and first principles calculations. The Hubbard model 
calculations of realistic models of disordered graphene show that 
only single-atom defects such as vacancies or chemical functionalizations 
unequally distributed over the two sublattices may result in a net 
magnetic moment. Further first principles calculations point out that 
the stacking order of graphite lies at the origin of ferromagnetism in 
proton bombarded graphite.

% Mean-field Hubbard model
To study the magnetic properties of disordered graphene I employ 
the mean-field approximation of the Hubbard model which proved to
describe carbon $\pi$-electron systems in good agreement with first principles calculations
\cite{Fernandez-Rossier07}. The corresponding Hamiltonian reads
\begin{equation}
	{\mathcal H} = - t \sum_{\langle i,j \rangle, \sigma} [ c_{i\sigma}^\dagger c_{j\sigma} + {\rm h.c.} ] + U \sum_{i, \sigma} n_{i\sigma} \langle n_{i -\sigma} \rangle, 
\end{equation}
where the first part is the single-orbital tight-binding Hamiltonian
while the second part accounts for the on-site Coulomb repulsion.
In this expression $c_{i\sigma}$ ($c_{i\sigma}^\dagger$) annihilates 
(creates) an electron with spin $\sigma$ at site $i$ and $\langle i,j \rangle$ 
stands for the pairs of bonded atoms. 
The expectation values of the spin-resolved density 
$n_{i\sigma}= c_{i\sigma}^\dagger c_{i\sigma}$ on atom $i$ are obtained 
from the eigenvectors of ${\mathcal H}$. The self-consistent solution of the problem 
provides the spin densities $M_i=(n_{i\uparrow}-n_{i\downarrow})/2$ on each 
atom $i$. 

The models of disordered graphene are generated by distributing randomly
point defects in a large $20 \times 20$ supercell (800 atoms).
The calculations are performed for 64 different random distributions 
in order to provide reliable statistical sampling.
This computational protocol yields the mean magnetic moments 
converged within $\sim$1\% both with respect to the supercell size 
and the number of configurations. 
Similar approach was successfully applied to study vacancy-induced magnetism
in oxide materials \cite{Bouzerar06}.
I consider the range of defect concentration $x = N_{\rm d}/N = 0.01-0.1$ with $N_{\rm d}$ out of $N$ atoms or bonds being affected. 
The values $U/t=1.0,1.33,1.67,2.0$ are studied
in order to investigate the role of on-site Coulomb repulsion magnitude.
While the value of hopping integral $t \approx 2.7$ eV
is well established for graphene, there is a growing debate regarding 
the on-site Coulomb repulsion $U$. It is worth pointing out that the magnitude of 
$U$ was inferred from magnetic resonance studies of neutral soliton states in 
{\it trans}-polyacetylene \cite{Su79}, a one-dimensional bipartite $sp^2$ carbon 
system closely related to graphene. The estimated $U = 3.0 - 3.5$~eV 
($U/t = 1.1-1.3$) \cite{Thomann85-Kuroda87} closely corresponds to 
$U/t=1.33$ considered here. 

The empirical value of $U/t$ can also be compared to the 
values obtained by establishing a relation between the results of first principles 
and mean-field Hubbard model calculations \cite{Pisani07}. The local density 
approximation (LDA) leads to $U/t \approx 0.9$ while the generalized gradient 
approximation (GGA) gives $U/t \approx 1.3$ in good agreement with the 
empirical value. It is somewhat cautionary that the results provided by generally more 
accurate hybrid density functionals correspond to $U/t \approx 2.0$ which is very close 
to the critical value $U/t \approx 2.2$ at which graphene accepts an antiferromagnetic
ground state \cite{Peres04}. Discrepancies between the results of hybrid density functional 
calculations and experimental data for {\it trans}-polyacetylene
have also been reported \cite{Bally00}. This urges to reconsider 
performance of these methods for describing magnetic $sp^2$ carbon systems. 

% Disorder classification
Disorder in graphene induced by chemical treatment or irradiation
with ions or electrons can be conveniently classified by the following three 
types of point defects. A carbon atom removed from the $sp^2$ lattice (i) 
(`vacancy`) corresponds to either true vacancy \cite{Lehtinen04,Yazyev07a} 
or the rehybridization into the $sp^3$ configuration. Such rehybridization
may result from chemical functionalization (e.g. hydrogen chemisorption 
\cite{Ruffieux00,Yazyev07a}) or binding of the interstitial carbon atoms 
in irradiated graphite \cite{Telling03,Yazyev07b}. 
A multiple-atom `vacancy` can be viewed as an ensemble of neighboring 
single-atom `vacancies`. In some cases a bivalent chemical functionalization 
results in breaking the bond between the pairs of adjacent atoms
without their rehybridization \cite{Lee06}. This leads to so-called 
bond dilution (ii). While (i) and (ii) give rise to undercoordinated 
carbon atoms, topological defects (iii) such as the Stone-Wales defect \cite{Stone86}
maintain the coordination number. This type of disorder
can be described as a permutation of connectivity involving two or more 
neighboring atoms. However, no magnetic moments due to the presence of the Stone-Wales 
defects were observed in this study as well as in
previous {\it ab initio} calculations \cite{Duplock04}. Below, 
I focus only on the magnetism induced by the `vacancies` and the 
bond dilution disorder. 

% Vacancies

\begin{figure}
\includegraphics[width=5.75cm]{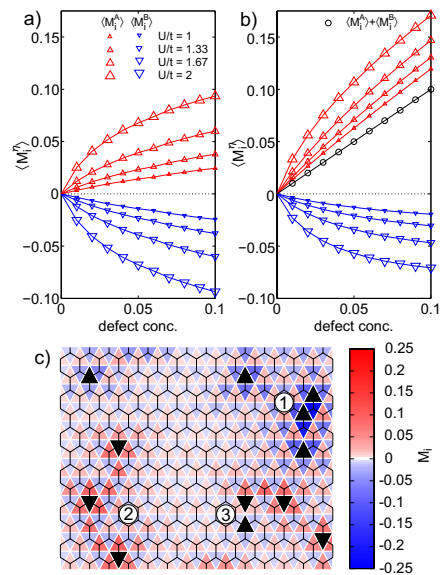}
\caption{\label{fig2} (Color online) Average magnetic moments of 
carbon atoms in $A$ and $B$ sublattices of graphene as a function
of `vacancy` concentration for different $U/t$ values. The `vacancies`
are either (a) distributed equally among sublattices or (b) 
belong to sublattice $B$ only. (c) Spin density distribution in 
a representative configuration with `vacancy` defects ($x=0.03$) 
in sublattices $A$ ($\blacktriangle$) and $B$ ($\blacktriangledown$).}
\end{figure}

The Hubbard model solutions for all configurations considered are 
characterized by the antiparallel orientation of magnetic moments 
on the two sublattices. Figs.~\ref{fig2}(a,b) show the sublattice-resolved 
average magnetic moments $\langle M_i^\eta \rangle$ ($\eta = A,B$) 
induced by varying concentrations of `vacancies`
for different values of $U/t$. For the case of defects equally 
distributed over the two sublattices [Fig.\ \ref{fig2}(a)] $\langle M_i^{\rm A} \rangle$ 
and $\langle M_i^{\rm B} \rangle$ compensate each other resulting in an
antiferromagnetic state with zero total magnetic moment 
$M = N(\langle M_i^{\rm A} \rangle + \langle M_i^{\rm B} \rangle)/2 =0$.
This result can be understood within the Lieb theorem \cite{Lieb89} 
which relates the total magnetic moment of a half-filled bipartite system to the difference of number of atoms in the two sublattices. 
In our case the two sublattices contain equal number 
of atoms resulting in zero net magnetic moment. 
The sublattice-resolved magnetic moments $\langle M_i^\eta \rangle \propto x^\gamma$ 
($0 < \gamma < 1$) show sub-linear dependence on $x$ and tend to increase 
with increasing the magnitude of $U/t$. In particular, for $U/t=1.33$ the  dependence on `vacancy` concentration $x$
can be accurately fitted by $|\langle M_i^\eta \rangle| \approx 0.234 \cdot x^{0.78}$. 
For the case of `vacancies` in sublattice $B$ only the total magnetic moment per unit
cell $M = x$ [open circles in Fig.\ \ref{fig2}(b)]. This is again in strict agreement
with the Lieb theorem predicting $2M=N_{\rm A}-N_{\rm B}$. The magnetic state originates 
from the Hund's rule population of $N_{\rm A}-N_{\rm B}$ zero-energy defect-induced quasilocalized 
states in sublattice $A$. However, these singly-occupied states induce exchange
polarization of the fully-occupied ones leading to a significant 
antiparallel magnetization of the atoms in sublattice $B$.
The induced magnetic moments follow an approximate dependence 
$\langle M_i^{\rm B} \rangle \approx -0.159 \cdot x^{0.69}$.

Figure\ \ref{fig2}(c) illustrates the distribution of magnetic
moments induced by the `vacancy` disorder in a representative
configuration ($x=0.03$) with defects in both sublattices. 
Magnetic moments develop in the nanometer-size regions characterized by the 
local domination of `vacancies` in one of the sublattices [{\bf 1} and {\bf 2} 
in Fig.\ \ref{fig2}(c)]. In these regions one can trace distinctive triangular patterns 
associated with individual quasilocalized states \cite{Mizes89-Kelly98}. The orientation of 
magnetic moments is determined by the antiferromagnetic coupling 
between the two sublattices. In contrast, the proximity of two 
defects pertaining to opposite sublattices [{\bf 3} in Fig.\ \ref{fig2}(c)] 
produces no local magnetic moments. This can be explained by
the fact that the quasilocalized states on different sublattices form a pair
of bonding/antibonding states \cite{Kumazaki07}, and the energy 
splitting increase with decreasing the distance between the two `vacancies`.
Such dispersion provides a mechanism for escaping the low-energy instability
which is preferred to the spin polarization. Increased contribution of this 
channel with increasing defect concentration explains the sublinear 
dependence of induced magnetic moments on $x$ for the case of `vacancies` 
in both sublattices. Both mechanisms contribute to stabilizing low-energy 
states. Indeed, band gaps increasing from 0.14~eV to 0.29~eV are found in the range of
concentrations $x=0.01 - 0.1$ and $U/t=1.33$. This agrees with
the recent experimental observation of insulating behavior 
in proton irradiated graphite \cite{Schindler08}.   

% Bond dilution

\begin{figure}
\includegraphics[width=5.75cm]{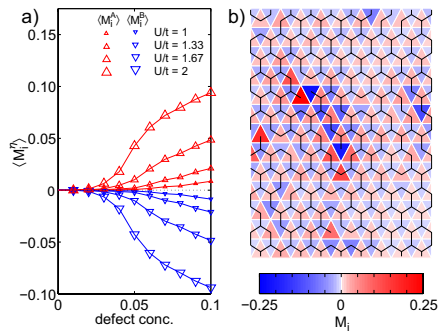}
\caption{\label{fig4} (Color online) 
(a) Average magnetic moments of carbon atoms in 
$A$ and $B$ sublattices of graphene as a function
of bond dilution for different $U/t$ values and
(b) spin density distribution in a representative 
configuration ($x=0.10$; $U/t=1.33$).}
\end{figure}

In contrast to the `vacancy` disorder, bond dilution does not 
change the number of atoms and, thus, 
can not produce a net magnetic moment. Nevertheless, antiferromagnetic
correlations develop as a result of bond dilution [Fig.\ \ref{fig4}(a)].
At low values of bond dilution ($x < 0.05$ for $U/t=1.33$) only very 
small average magnetic moments develop as a results of occasional 
occurrence of highly undercoordinated atoms. For larger values of $x$
spin polarization occurs in extended regions of higher local concentration
of broken bonds as shown in Fig.\ \ref{fig4}(b) for $x=0.1$ and $U/t=1.33$.
Such cumulative behavior is similar to the quantum phase transition 
predicted for hexagonal graphene nanoislands \cite{Fernandez-Rossier07}.

\begin{figure}
\includegraphics[width=6.5cm]{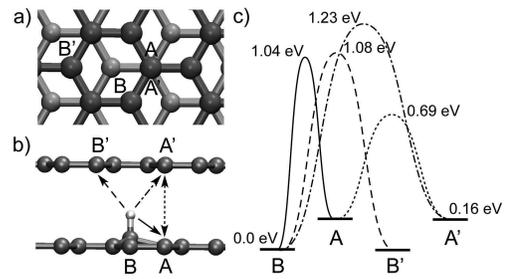}
\caption{\label{fig5} (a) Inequivalent carbon atoms ($A$ and $B$) in 
graphite. (b) Possible pathways for the diffusion 
of chemisorbed hydrogen in the $ab$ plane of graphite. (c) Relative 
energies of the potential energy surface minima and transition states
for the diffusion of chemisorbed hydrogen in graphite.}
\end{figure}

% Proton-induced 
One can now conclude that a macroscopic magnetic moment
may appear only in the presence of `vacancy` disorder unequally 
distributed over the two graphene sublattices. In bulk
graphite and multilayer graphene the sublattices are 
inequivalent due to the stacking order [Fig.\ \ref{fig5}(a)]. 
Irradiation of graphite with high-energy ions initially produces 
(i) vacancy-interstitial pairs due to knock-on recoils and 
(ii) chemisorbed ions stopped by the bulk material. It was found
that proton bombardment of graphite leads to much stronger magnetic 
signal than the He-ion irradiation where no chemisorption takes place. 
This suggests that the hydrogen chemisortion plays an important role in 
the onset of magnetism in proton bombarded graphite. 

The effect of stacking order in graphite is further investigated by means of 
first principles calculations. I considered a $4 \times 4 \times 1$
supercell of 
ABA stacking graphite within the GGA density functional theory framework \cite{Perdew96, QE}. 
Chemisorption of a proton in position $B$
was found to be 0.16~eV lower in energy than in position $A$ [Fig.\ \ref{fig5}(b,c)].
This is explained by the steric repulsion between
the proton chemisorbed on atom $A$ and carbon atom $A'$ of the adjacent graphene layer. 
A magnetic moment of 1~$\mu_B$ localized in the functionalized 
graphene sheet is observed for the hydrogen chemisorption in both inequivalent positions of the crystalline lattice of graphite.
This is in full agreement with the previous {\it ab initio} 
calculations \cite{Yazyev07a,Boukhvalov08} and the present Hubbard model results for graphene.  
The barrier for the migration of proton initially trapped in position 
$A$ to the neighboring position $B$ is 0.88~eV (path $A \rightarrow B$ in 
Fig.\ \ref{fig5}(c)). Such migration can take place at the experimental
conditions. The overall barrier for the in-plane 
diffusion of protons in graphite (paths $A \rightarrow B$,  
$B \rightarrow B'$ and $A' \rightarrow B$ in Fig.\ \ref{fig5}(b,c)) defined by the lowest 
barrier of path $A \rightarrow B$ is 1.04~eV. Diffusion in the direction 
of $c$-axis requires much higher activation energy of $\sim$4~eV. Free diffusion 
of protons in graphite will result 
in either saturation of vacancy dangling bonds \cite{Lehtinen04} or in 
the formation of energetically favorable aggregates with both 
sublattices being populated \cite{Boukhvalov08}. Magnetic moments of the hydrogen
aggregates quench due to the large splitting of quasilocalized states in the 
complementary sublattices. In addition, the condition of preserving the 
stacking order of graphite limits the maximum irradiation dose and intensity. Subsequently,
there exist only relatively narrow range of conditions at the 
sublattice-discriminating hydrogen chemisorption can be achieved. This accords with the fact 
magnetic moments are induced only in the narrow ring surrounding the irradiated
spot \cite{Ohldag07}. 
% suggsestion

% Recombination
Similarly to hydrogen chemisorption, single-atom vacancies produced by 
the knock-on recoils induce low-energy quasilocalized states \cite{Yazyev07a}. 
In contrast, interstitial carbon atoms in the energetically preferred
configuration \cite{Telling03,Li05} lead to the rehybridization of the pairs
of neighboring atoms in the adjacent graphene layers. This situation is 
equivalent to the case of di-vacancies providing no low-energy states
required for developing magnetic moments. Cross-sections for the momentum 
transfer due to knock-on collisions are likely to be equal for both
$A$ and $B$ carbon atoms in graphite. However, the stacking order of 
graphite has strong influence on the kinetics of interstitial-vacancy 
recombination. Instantaneous recombination of low-energy recoils 
was found to be significantly more likely for the atoms in 
position $A$ \cite{Yazyev07b}. At longer time scales, diffusing interstitials 
encounter different barriers for the recombination with 
vacancies in these two positions. First principles calculations show 
different minimum transition barriers of 0.04~eV and 0.22~eV for $A$ and $B$
vacancies, respectively. Both values are below
the diffusion barrier of interstitial atoms ($\sim$0.5~eV \cite{Ma07}) suggesting that only
very small difference in populations of the two sublattices by vacancy defects 
can be achieved in practice. Nevertheless, both vacancies and chemisorbed hydrogen 
atoms are found to populate preferably sublattice $B$ thus producing ferrimagnetic 
order with the spin polarization larger for sublattice $A$.

% Conclusions
In conclusion, I have shown the crucial role of single-atom defects
in combination with a sublattice-discriminating mechanism for developing
ferromagnetic order in graphene-based materials. In graphite the role of 
such mechanism is played by the stacking order of graphene layers. 
This suggestion is confirmed by the {\it ab initio} investigation
of hydrogen chemisorption and vacancy defects in proton 
bombarded graphite. The results of this study pave a way for tailoring
carbon based magnetic materials by means of irradiation and chemical treatment of graphite and other graphene derivatives.

I acknowledge the comments of P.~Esquinazi, L.~Helm, A.~Krasheninnikov and W.~L.~Wang.
The work was partly supported by the Swiss NSF. First principles calculations
have been performed at the CSCS (Manno).

 %\bibliography{disorder}

\end{document}